\documentclass[a4paper,10pt]{article}
\usepackage{graphicx}
\usepackage{cite}

\def\half{\frac{1}{2}}
\def\nn{\nonumber}

\begin{document}

\title{Static spherically symmetric solutions\\ of Einstein field equations with\\ radial dark matter}
\author{Igor Nikitin\\Department of High Performance Analytics\\Fraunhofer Institute for Algorithms and Scientific Computing\\Schloss Birlinghoven, 53757 Sankt Augustin, Germany\\ \\igor.nikitin@scai.fraunhofer.de}
\date{}

\maketitle

\begin{abstract}
We study a static spherically symmetric problem with a black hole and radially directed geodesic flows of dark matter. The obtained solutions have the following properties. At large distances, the gravitational field produces constant velocities of circular motion, i.e., flat rotation curves. At smaller distances, the field switches to Newtonian regime, then to Schwarzschild regime. Deviations from Schwarzschild regime start below the gravitational radius. The dark matter prevents the creation of event horizon, instead, a spherical region possessing extremely large redshift is created. The structure of space-time for the obtained solutions is investigated and the implications for the models of the galaxies are discussed.
\end{abstract}

\begin{figure}[h]
\centering
\includegraphics[width=0.8\textwidth]{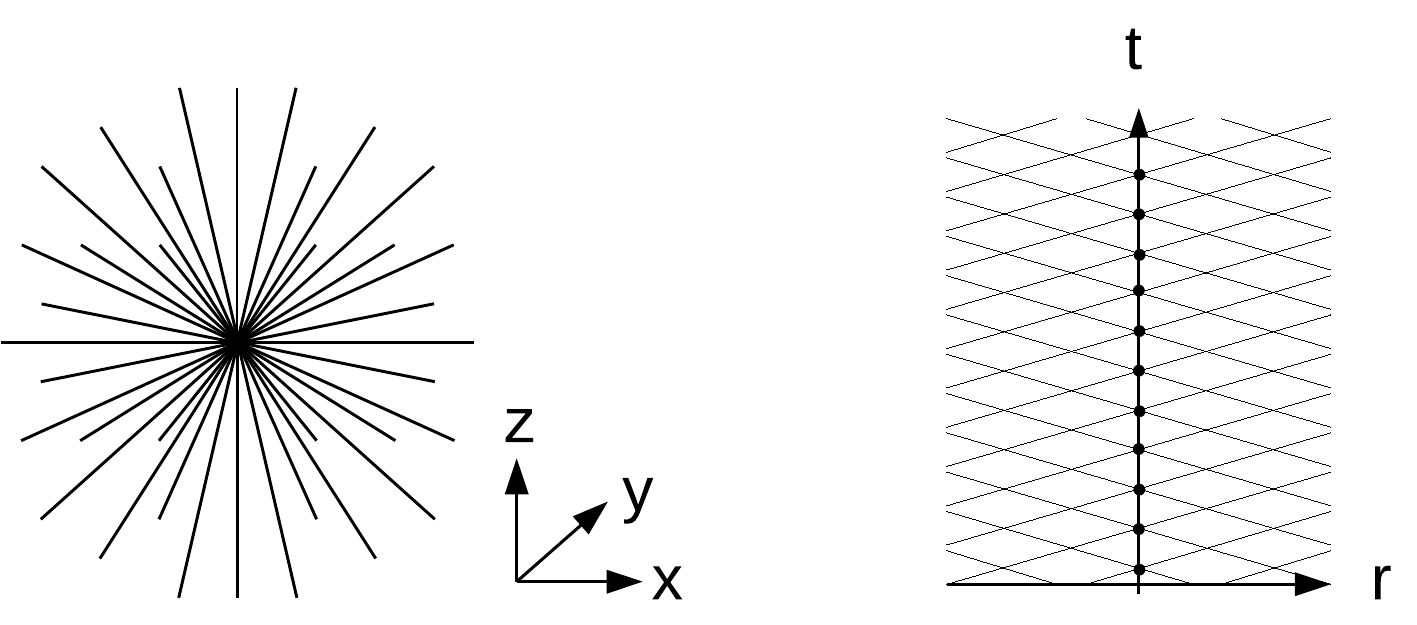}
\caption{Static spherically symmetric problem with radially diverging flows of dark matter. On the left: spatial projection, on the right: projection involving time $t$ and radial coordinate $r$. The pattern on the right is composed of two overlayed flows in ingoing and outgoing directions.}
\label{f1}
\end{figure}

\section{Introduction}\label{intro}
It has been shown in \cite{tachyo-dm} that asymptotically flat rotation curves, typically observed in spiral galaxies, can be reproduced by prescribing dark matter distribution of a special type. The dark matter is composed of non-interacting particles, moving along geodesic world lines, radially diverging from the center of the galaxy. In a static spherically symmetric scenario, a spatial projection of this configuration, being viewed from a large distance, looks like Fig.\ref{f1} left. The projection involving time and radial coordinate, shown on Fig.\ref{f1} right, comprises ingoing and outgoing radial flows of the dark matter. Stationarity of solution implies a necessary condition of energetic balance, the flow of energy through spheres $r=Const$ for ingoing and outgoing flows must coincide. Equivalently, the total flow of energy must vanish, implying a condition $T_{tr}=0$ on the corresponding component of energy-momentum tensor. Solving this scenario in weak field approximation, \cite{tachyo-dm} obtains the rotation curves composed of a positive constant and short-range Newtonian term. The purpose of the present work is to continue this consideration in the range of strong fields.

Concerning the composition, the study in \cite{tachyo-dm} was concentrated on tachyonic dark matter. The tachyonic models of dark matter are becoming widespread \cite{Shiu,Frolov,Bagla,Arefxeva,Davis}. Considering a scenario proposed in \cite{tachyo-dm}, let us introduce a supermassive elementary particle $X$, decaying into two particles of normal and tachyonic type: $X\to n+n$, $X\to t+t$. Considering both vertices in the rest frame of $X$, it is easy to verify that the both processes satisfy the conservation laws and are kinematically allowed. Joining these processes together: $n+n\to X\to t+t$, one obtains a process of conversion of normal matter to tachyons. Due to high mass threshold, this process is suppressed at low energies and becomes active only at the energies comparable with the mass of $X$. Singularities of the black holes are the places where arbitrarily large energies are available, therefore they can become generators of tachyons. The event horizons surrounding the singularities are transparent for superluminal particles, thus the tachyons can leave the black holes and further can propagate along radial geodesics. At low energies they do not enter in direct interaction with normal matter and reveal their presence only by gravitational effects, i.e., possess the properties of astrophysical dark matter. Radially diverging configuration of world lines leads to the density decreasing as $\rho\sim r^{-2}$, exactly what is necessary for reproduction of flat rotation curves.

Strictly speaking, this consideration does not rule out the dark matter of normal type, consisting of massive or null particles. The condition of energetic balance can be satisfied for the flows of normal matter and for mixed flows of normal and tachyonic matter. In all these cases the computation \cite{tachyo-dm} produces asymptotically flat rotation curves. 
We will show in this paper that such indifference to the type of matter persists also in strong fields. The argument that tachyons go out of event horizons (while normal matter does not) can be countered by the fact that the considered solutions do not have event horizons. For the purpose of objectivity, we will consider all types of matter in this paper, making the distinctions among them when necessary.

We will consider a static scenario, where the field configuration and ingoing and outgoing flows do not change in time. It differs from a dynamical collapse scenario, when the ingoing normal matter is once converted to outgoing tachyonic matter, which then disperses. We think about the static solution as an asymptotic state, a remnant of a real collapse, which possesses persistent ingoing and outgoing flows and lasts indefinitely. In this paper we will not consider a dynamical simulation of real collapse, which can end in such self-sustaining state. This can be an interesting topic for future work.

There are also several complex and interesting questions concerning the relation of causality principle with $T$-invariance of the given problem. In full generality these questions were considered in the classical works by Wheeler and Feynman \cite{wf1,wf2}, later in \cite{gold,ft,price} and recently in \cite{cosmic-web}. Following argumentation in these papers, the direction of the world lines as well as the direction of energy-momentum vector is a matter of convention. The action of the particle, equal to the length of the world line, is invariant under reversal of its direction, as a result, all physically meaningful quantities as tensor of energy-momentum, are invariant under such reversal. The ingoing flows of matter are equivalent to the outgoing flows directed backward in time. The energetically balanced processes are equivalent to the self-compensated emission of positive energy into the future and negative energy into the past. 

The tachyons are kinematically different from the normal particles, in the sense that tachyon momentum is constrained by a mass shell condition $p_0^2-\vec p^2=-m^2$, defining one-sheeted hyperboloid, while the normal matter corresponds to $p_0^2-\vec p^2=m^2$, two-sheeted hyperboloid, from which only the sheet $p_0>0$ is normally taken. Therefore continuous Lorentz transformations can make tachyons reversing time direction, while the momenta of normal particles remain future-directed. Nevertheless, one can formally introduce past-directed normal particles, e.g., by transferring the reagents in particle reactions to one side, from $X\to n+n$ to $0\to-X+n+n$. If one considers this vertex as a network of three world lines connected in one node, the last record means that all energy-momenta vectors directed away from the node are summed to zero, a condition coincident with the stationarity of action for this network \cite{cosmic-web}. In this view, there is not much difference between tachyonic and normal particles, both can be considered as future-directed or past-directed, while the physical content of the theory is completely invariant under the reversal of this direction. 

The interpretation involving the concepts of causality and entropy is not so invariant. For example, it is easy to explain the appearance of the radially outgoing flow of matter by its creation in the center. In $T$-symmetric theories, it is equally easy to explain the radially ingoing flow of matter by $T$-conjugation. However, it is difficult to explain the radially ingoing flow in terms of usual causal and entropic concepts. It is not clear why the ingoing flow must be directed to the center already at very large distances, where the gravitational influence of the center is negligible and there are no forces favoring this direction. The obstacle here can be a methodological asymmetry in the treatment of initial and final conditions mentioned in \cite{cosmic-web}. While the causal reasoning is based on setting initial conditions and tracking the evolution into the future, $T$-symmetric problems can involve final conditions, e.g., a point in space-time where the world line must end. In everyday experience such conditions are quite unusual, this leads to the difficulties with their interpretation. Further we will rely on $T$-symmetry of the problem and for the discussion of causality questions forward the interested reader to the references above.

In this paper we will study the configuration shown on Fig.\ref{f1}, with two layers, formed by $T$-conjugated ingoing and outgoing flows. This configuration automatically satisfies energetic balance condition. We note, however, that energetic balance requires neither that there are only two layers nor that they are $T$-conjugated. In general case there can be many or even a continuous set of layers, not possessing $T$-symmetry, but summed to zero in energetic balance condition. Here we restrict ourselves to the study of a particular set of solutions. 

In Sec.~\ref{sec_efe} we formulate the field equations, defining static spherically symmetric problem with the dark matter of radial type. The equations will be transformed to a system of two ordinary differential equations, which will be solved numerically for a representative set of model parameters. In Sec.~\ref{sec_geode} we describe a structure of geodesics, defining the motion of probe particles in the obtained gravitational field. The radial and circular geodesics will be investigated, with the related questions of geodesic reachability and stability of circular orbits. In Sec.~\ref{sec_galaxy} we match the model with parameters of the Milky Way galaxy, describing the ranges, transitions and associated physically observable effects. We will also discuss possible extensions of the model.

\section{Solving field equations}\label{sec_efe}

\begin{figure}
\centering
\includegraphics[width=0.49\textwidth]{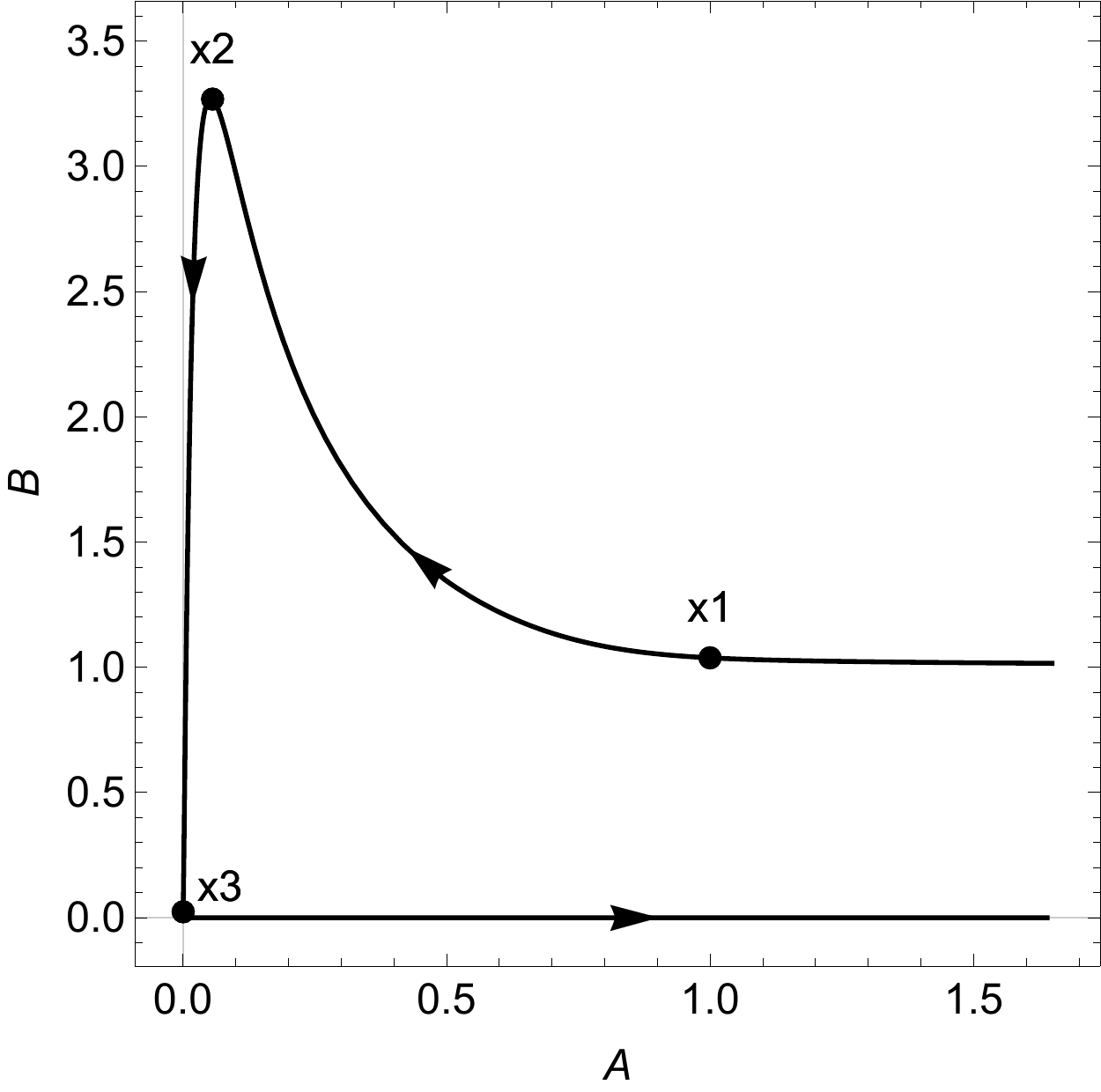}
\includegraphics[width=0.49\textwidth]{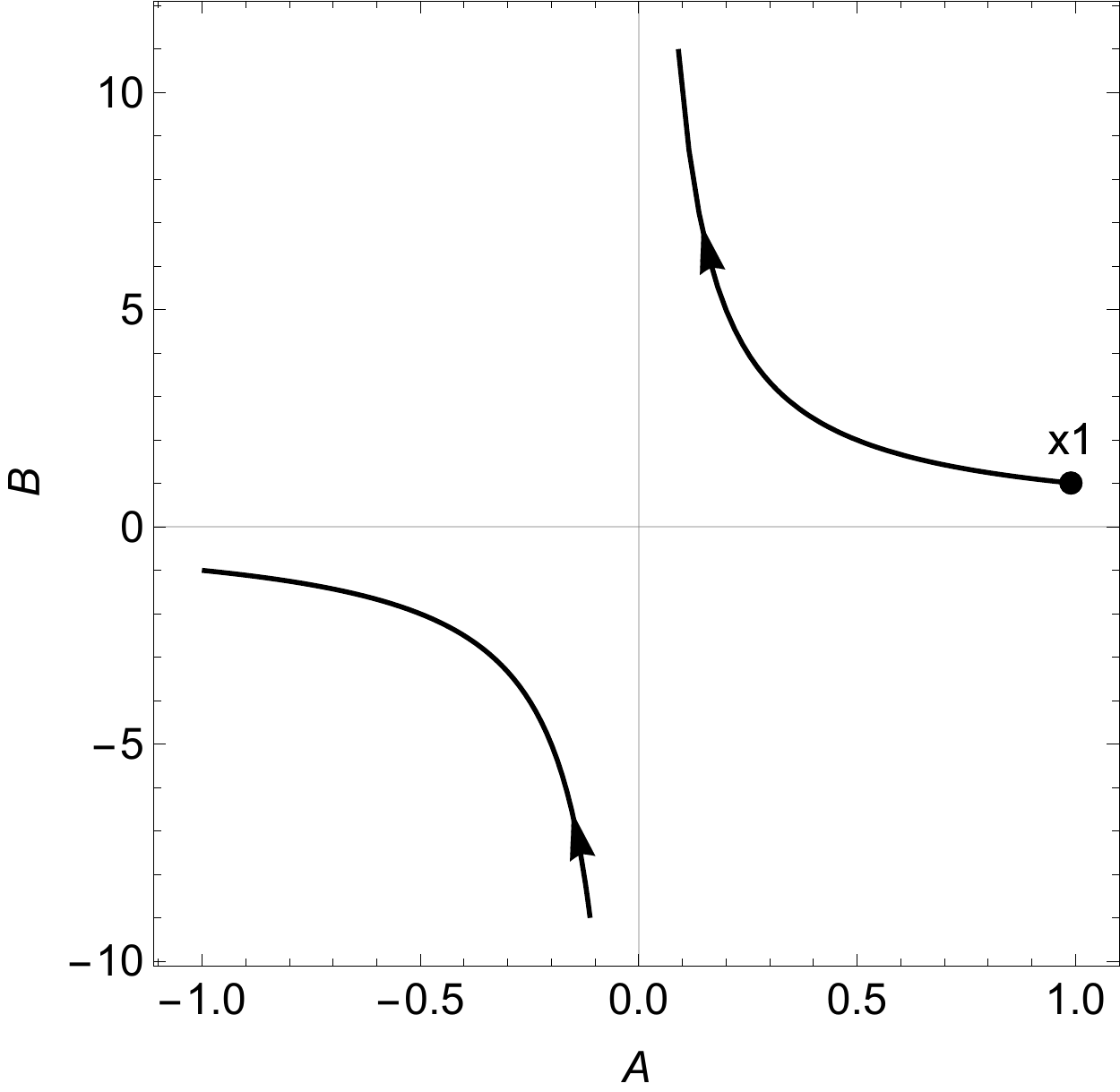}
\caption{The evolution of functions $A,B$, defining the metric of space-time, with variation of $r$. On the left: solution of RDM model with intensity factor $\epsilon=0.04$. On the right: Schwarzschild solution ($\epsilon=0$). For Schwarzschild solution $A\to+0$ and $B\to+\infty$ while $r\to r_s+0$, then $A,B$ become negative. This singularity indicates the presence of an event horizon in the solution. For RDM solution the evolution of $A$,$B$ is constrained to the positive region, there is no event horizon.}
\label{f2}
\end{figure}

We are set to solve a system of Einstein field equations coupled with geodesic flow equations, in standard denotations \cite{blau}:
\begin{eqnarray}
&&R^{\mu\nu}-\half g^{\mu\nu}R=8\pi G_N T^{\mu\nu},\label{eq_efe}\\
&&u^\nu\nabla_\nu u^\mu=0,\ \nabla_\mu\rho u^\mu=0.\label{eq_gfe}
\end{eqnarray}
We fix a system of units $c=1$, $4\pi G_N=1$. 
Using spherical coordinates $(t,r,\theta,\phi)$, standard line element for static spherically symmetric space-time:
\begin{equation}
ds^2=-A(r)dt^2+B(r)dr^2+r^2(d\theta^2+\sin^2\theta\; d\phi^2)
\end{equation}
and energy-momentum tensor of the form:
\begin{equation}
T^{\mu\nu}=\rho(r)(u_+^\mu(r)u_+^\nu(r)+u_-^\mu(r)u_-^\nu(r)),
\end{equation}
where $\rho(r)$ is a density and $u_\pm(r)=(\pm u^t(r),u^r(r),0,0)$ are velocity vectors of $T$-symmetric outgoing and ingoing radial dark matter flows, obtain from (\ref{eq_gfe}):
\begin{eqnarray}
&&\rho'u^r  +  (4/r + A'/A + B'/B)\rho u^r/2 + \rho(u^r)'=0,\\
&&(u^t A'/A + (u^t)')u^r=0,\\
&&(u^t)^2 A' + (u^r)^2 B' +  2 B u^r (u^r)'=0.
\end{eqnarray}
A solution $u^r=0$ leads to flat space-time and zero density of dark matter, while non-trivial solution has a form: 
\begin{eqnarray}
&&\rho=c_1/\left(r^2u^r\sqrt{AB}\right),\ u^t=c_2/A, \ u^r=\sqrt{c_2^2+c_3A}/\sqrt{AB},\label{eq_geode}
\end{eqnarray}
with positive constants $c_{1,2}$. The third constant defines a covariant norm $c_3=u_\mu u^\mu$ and can be set to three discrete values:
\begin{itemize}
\item $c_3=1$, tachyonic radial dark matter (TRDM),
\item $c_3=0$, null radial dark matter (NRDM),
\item $c_3=-1$, massive radial dark matter (MRDM).
\end{itemize}
The system (\ref{eq_efe}) then reduces to:
\begin{eqnarray}
&&rA'=-A+AB+ 4c_1B\sqrt{c_2^2 + c_3A},\\
&&rB'=B/A\left(A-AB+4c_1c_2^2B/\sqrt{c_2^2 + c_3A}\right).
\end{eqnarray}
In particular case $c_1=0$ we obtain vacuum equations with well known solution:
\begin{eqnarray}
&&A=1-r_s/r,\ B=(1-r_s/r)^{-1},
\end{eqnarray}
where $r_s$ is Schwarzschild radius. Further, introducing transformation to a new variable and new constants
\begin{eqnarray}
&&x=\log r,\ c_4=4c_1c_2,\ c_5=c_3/c_2^2,\label{eq_c45}
\end{eqnarray}
the system can be rewritten to:
\begin{eqnarray}
&&dA/dx=-A+AB+ c_4B\sqrt{1 + c_5A},\label{eq_dAdx}\\
&&dB/dx=B/A\left(A-AB+c_4B/\sqrt{1 + c_5A}\right).\label{eq_dBdx}
\end{eqnarray}
We solve this system numerically and show the result on Fig.\ref{f2}, in comparison with Schwarzschild solution.

The first difference is that in Schwarzschild solution $A$ goes to zero and $B$ tends to infinity while approaching $r_s$, then $A,B$ become negative. In this transition the signatures of $r$ and $t$ in the metric are interchanged, leading to the build-up of an event horizon and other phenomena associated with the black hole solution. For RDM solution at some point the function $B$ stops to grow. The reason is that for small $A$ the dark matter $c_4$-term starts to dominate and breaks the tendency. After that point both $A$ and $B$ fall down to nearly zero values. None of them crosses zero, however. While $B$ continues tending to zero, at a certain point $A$ stops to fall. The reason is that the first $(-A)$ term in (\ref{eq_dAdx}) starts to dominate, leading to further increase of $A$ with decrease of $x$.

\begin{figure}
\centering
\includegraphics[width=\textwidth]{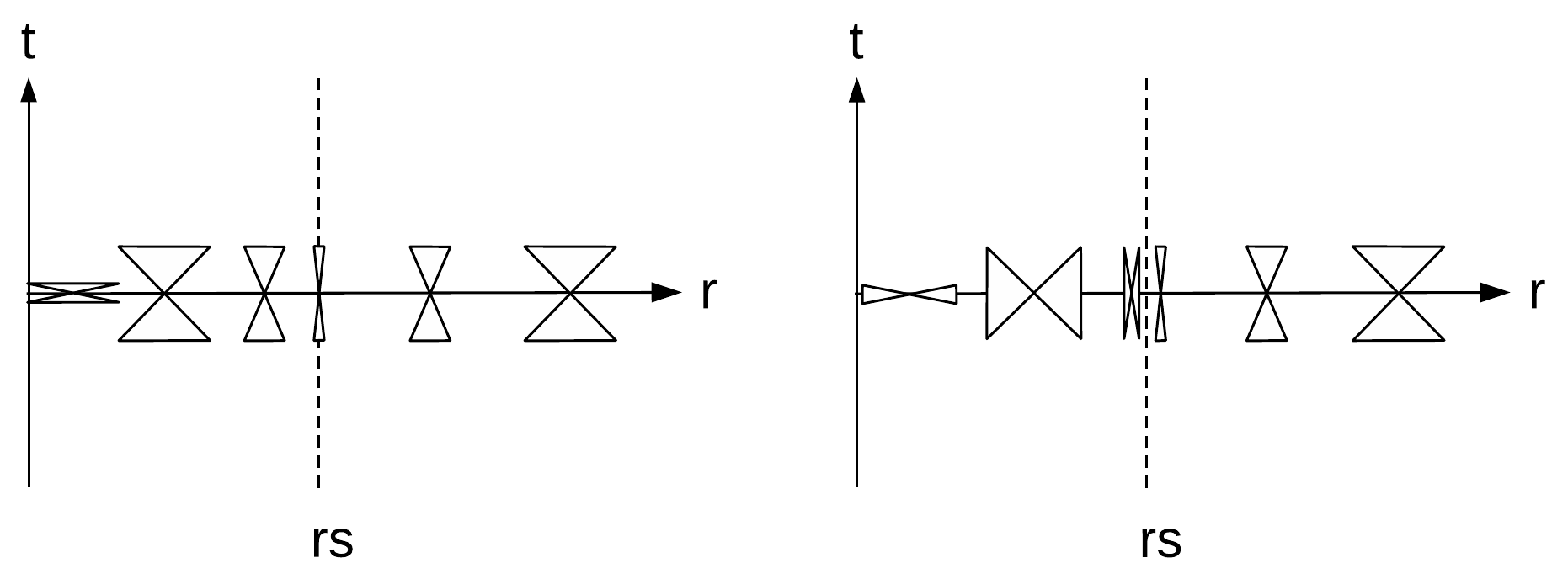}
\caption{The distribution of light cones, on the left: RDM solution, on the right: Schwarzschild solution. Time-like directions correspond to the closed sides of the cones. For Schwarzschild solution the cones flip their direction while passing $r_s$. This, again, indicates the event horizon, at $r<r_s$ time-like world lines point towards singularity $r=0$. For RDM solutions the cones do not flip and the horizon is absent.}
\label{f3}
\end{figure}

Thus, we see that the dark matter term in the given model works as a barrier, repelling the solution from the region of negative $A,B$, in this sense preventing the creation of an event horizon. To illustrate the related differences of causal structure, on Fig.\ref{f3} we have shown the distribution of light cones in RDM and Schwarzschild solutions. The closed sides of the cones correspond to time-like directions. For Schwarzschild solution the cones become infinitely thin approaching $r_s$ and further flip their direction. Under the horizon, time-like vectors are directed towards the center and the time-like world lines end on the singularity. For RDM the cones have non-zero thickness intersecting $r_s$ and further preserve their direction. The time-like observer can go under $r_s$ and then return back.

\begin{figure}
\centering
\includegraphics[width=0.49\textwidth]{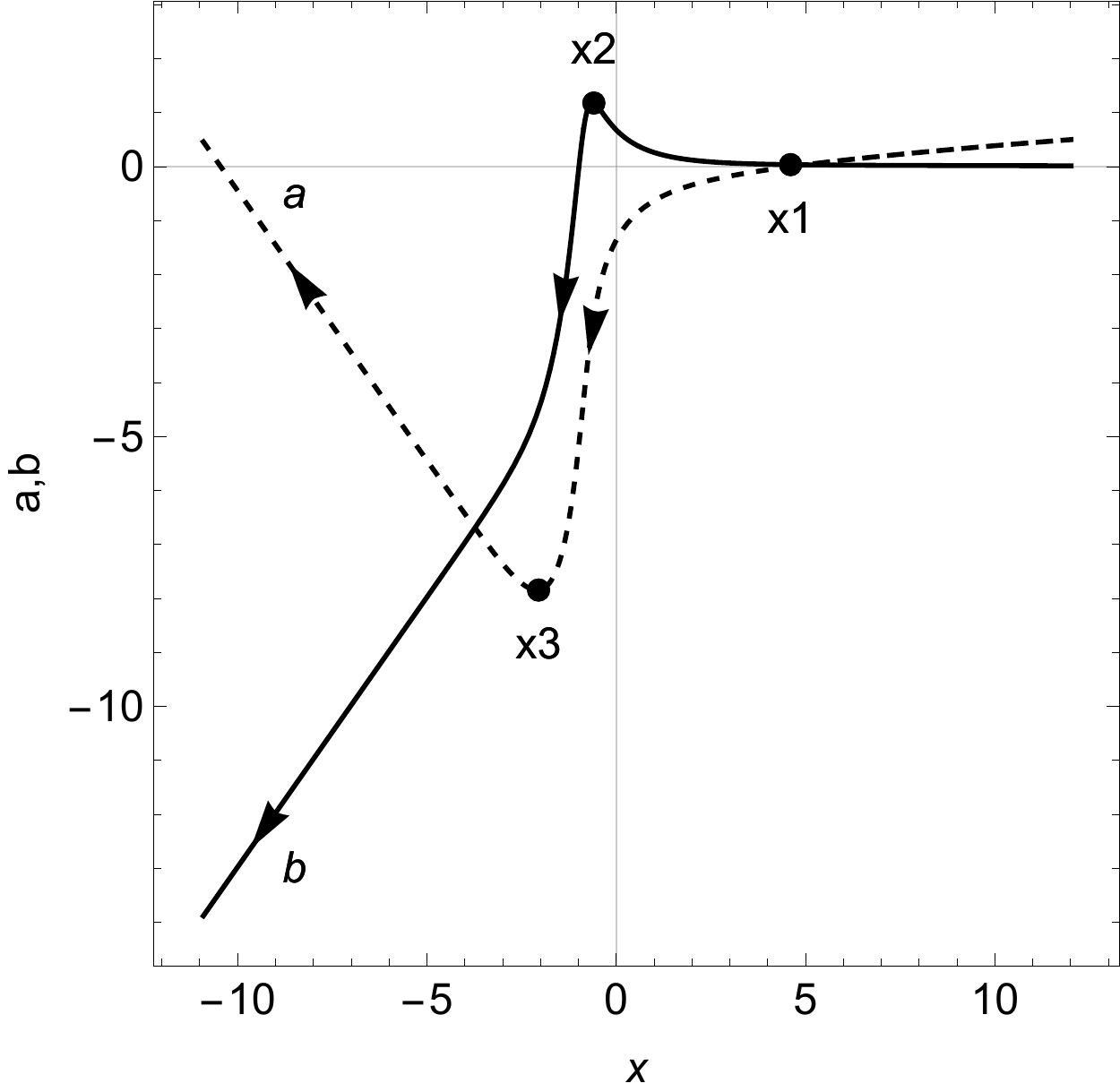}
\includegraphics[width=0.49\textwidth]{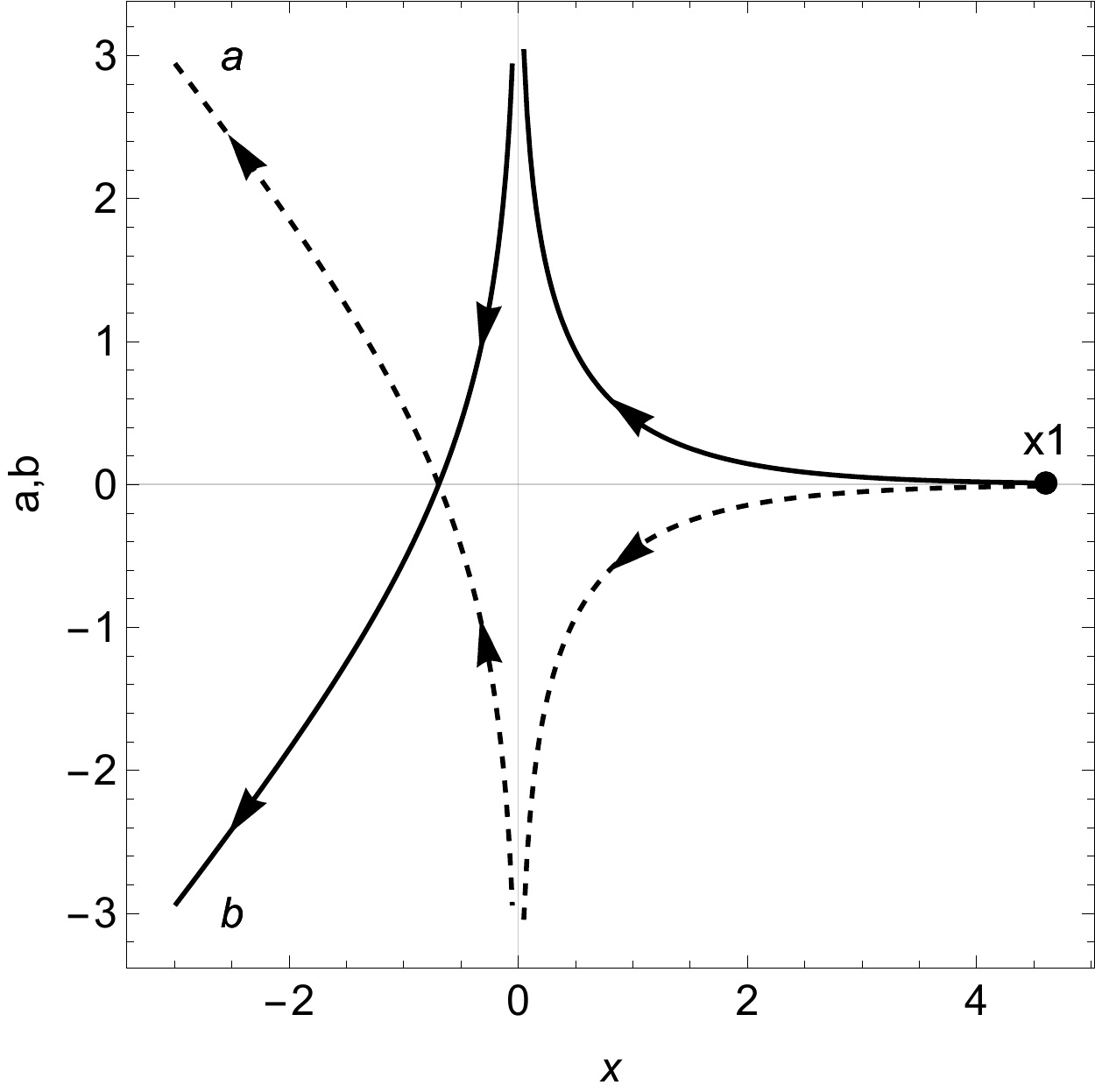}
\caption{The evolution of $a=\log|A|$, $b=\log|B|$ as functions of $x=\log r$. On the left: RDM solution ($\epsilon=0.04$), on the right: Schwarzschild solution, $x_1$ is a starting point of integration. For RDM solution $b$ reaches a maximum in the point $x_2$, then the functions $a$ and $b$ rapidly fall down. The fall stops when $a$ reaches a minimum in the point $x_3$, then the functions symmetrically go apart. For Schwarzschild solution the functions are symmetric over the whole range and go to infinity at $r\to r_s$.
}
\label{f4}
\end{figure}

Since $A,B$ for RDM solution remain in positive region, we can apply a transformation $a=\log A$, $b=\log B$ and once more rewrite the equations:
\begin{eqnarray}
&&da/dx=-1+e^b+c_4e^{b-a}\sqrt{1 + c_5e^a},\label{eq_dadx}\\
&&db/dx=1-e^b+c_4e^{b-a}/\sqrt{1 + c_5e^a}.\label{eq_dbdx}
\end{eqnarray}
On Fig.\ref{f4} we compare these functions with the similar functions $a=\log|A|$, $b=\log|B|$ for Schwarzschild solution. We set $r_s=1$ for both solutions and start from a point $x_1$ located well above $r_s$. On RDM solution $b$ reaches a maximum in a point $x_2$ located right under $r_s$. After that the profiles fall down in almost parallel way. In a point $x_3$ $a$-profile reaches a minimum and the profiles go apart with opposite slopes. Schwarzschild solution has similar profiles on initial and final stage. The differences are: infinities of Schwarzschild solution on the horizon and an intermediate slide of RDM profiles to the negative region. 

\paragraph*{Regimes and transitions.} In initial point $x_1$ we set $a_1=0$, corresponding to $A_1=1$. Equivalently, time variable $t$ is set to a proper time of static observer in the point $x_1$. Being far from $r_s$, we also set $b_1$ close to zero, nearly flat space-time in this point. In this regime the equations can be linearized:
\begin{eqnarray}
&&da/dx=b+c_6,\ db/dx=-b+c_7,\label{eq_lin}\\
&&c_6=c_4\sqrt{1 + c_5},\ c_7=c_4/\sqrt{1 + c_5},\label{eq_c67}
\end{eqnarray}
with the solution
\begin{eqnarray}
&&a=Const + (c_6+c_7) x - r_se^{-x},\ b=c_7 + r_se^{-x}.\label{eq_z1}
\end{eqnarray}
The solution coincides with that in \cite{tachyo-dm}:
\begin{eqnarray}
&&a=Const+2\epsilon\log r -r_s/r,\ b=2C_1+r_s/r
\end{eqnarray}
after the following matching of the constants:
\begin{eqnarray}
&&\epsilon=(c_6+c_7)/2,\ C_1=c_7/2,
\end{eqnarray}
while the additive constant to $a$ in both cases is arbitrary, equivalent to rescaling of time variable. The equations in \cite{tachyo-dm} were derived in the frames of weak field approximation, here we confirm this result by the independent computation. The function 
\begin{eqnarray}
&&\varphi=a/2=Const+\epsilon\log r -r_s/(2r)\label{eq_phi}
\end{eqnarray}
for weak fields defines gravitational potential and its derivative
\begin{eqnarray}
&&a_r=v^2/r=\epsilon/r +r_s/(2r^2)\label{eq_ar}
\end{eqnarray}
gives acceleration on circular orbits. At large $r$ the dark matter related term $\epsilon/r$ dominates over Newtonian term $r_s/(2r^2)$, producing asymptotically flat rotation curves $v^2=\epsilon$ for the models of spiral galaxies. This makes $\epsilon$ a measurable parameter, e.g., for the Milky Way $v\sim200$km/s, $v/c\sim6.6\cdot10^{-4}$ and $\epsilon\sim4\cdot10^{-7}$. In this view, it is convenient to take $(\epsilon,r_s,c_2,c_3)$ as independent model parameters, then reexpress
\begin{eqnarray}
&&c_1=\epsilon\sqrt{c_2^2 + c_3}/(4c_2^2 + 2c_3),
\end{eqnarray}
and use (\ref{eq_c45}), (\ref{eq_c67}) to define $c_{4-7}$. In numerical integration the starting point can be set to
\begin{eqnarray}
&&x_1=\log r_1,\ a_1=0,\ b_1=c_7 + r_se^{-x_1}. 
\end{eqnarray}
Considering further (\ref{eq_phi}) in the range of weak field approximation, the dark matter term in the acceleration prevails over Newtonian term at $r>r_{1a}=r_s/(2\epsilon)$, while at $r<r_{1a}$ Newtonian term prevails. Further, non-Newtonian effects appear, as a switching point one can select the radius of innermost stable circular orbit (ISCO) $r_{1b}\sim3r_s$. In this regime the gravitational field is close to Schwarzschild solution with dominating second term in (\ref{eq_dadx}),(\ref{eq_dbdx}).

In the next switching point $x_2$ the third term in (\ref{eq_dadx}),(\ref{eq_dbdx}) starts to dominate. Characteristic value is $a_2\sim\log c_4$. Here $\log c_4\sim\log\epsilon$, at small $\epsilon$ and fixed $c_{2,3}$. For Schwarzschild solution at this point $b_2=-a_2$, while for RDM solution the influence of the third term in (\ref{eq_dbdx}) makes $b$-profile going deeper, $b_2=-a_2-c_8$, with a positive constant $c_8$. For Fig.\ref{f4} the value $c_8=2.122$.   
The third term in (\ref{eq_dadx}),(\ref{eq_dbdx}) dominates further in the range between $x_2$ and $x_3$. Preserving only this term in the equation and assuming $a$ so negative that $|c_5e^a|\ll1$, the equations are reduced to:
\begin{eqnarray}
&&da/dx=db/dx=c_4e^{b-a},
\end{eqnarray}
with the solution
\begin{eqnarray}
&&a=c_4e^{c_9}x+c_{10},\ b=a+c_9,\label{eq_z2}
\end{eqnarray}
where $c_{9,10}$ are new integration constants, whose values can be found in connection of profiles (\ref{eq_z1}) and (\ref{eq_z2}) in the point $x_2$. For small $\epsilon$ the value $c_9$ is positive and large, since $b_2\gg a_2$ near the horizon. Matching of the constants in point $x_2$ gives a rough estimation $c_9=-2\log c_4-c_8$. Therefore, in the considered range $a$ and $b$ are rapidly falling down with decreasing $x$, possessing equal slopes. 

At a finer level, computing a difference and a ratio of (\ref{eq_dadx}),(\ref{eq_dbdx}) under the same assumptions:
\begin{eqnarray}
&&d(b-a)/dx=2-2e^b,\ d(b-a)/db=(2-2e^b)/(c_4e^{b-a}),\label{eq_ba}
\end{eqnarray}
we see that on the way of $a,b$ down the difference $b-a$ is gradually reduced. Finally, at $b_3-a_3\sim-\log c_4$ the third term in (\ref{eq_dadx}),(\ref{eq_dbdx}) equalizes the first, the constant one, leading to a new switch. To estimate its position, we integrate the last equation in (\ref{eq_ba}):
\begin{eqnarray}
&&d(c_4e^{b-a})=d(2b-2e^b),
\end{eqnarray}
and after substitution of the boundary values for points $x_2$ and $x_3$ have:
\begin{eqnarray}
&&b_3\sim-1.5e^{-c_8}/c_4,\ a_3\sim b_3+\log c_4,\ a_3,b_3\sim-c_{11}/\epsilon.
\end{eqnarray}
The new constant $c_{11}$ is positive and can be estimated collecting the previous constants together. We note, however, that the matching of the curves in the boundary points can serve only as a rough estimation for this constant. Preserving only one leading term in the equation gives insufficient precision near the boundary points, where the regime change is happening. Although this estimation allows to extract the main $1/\epsilon$-dependence, the constants should be better defined by numerical integration. On the other hand, the constant $c_{11}$ seems to be insensitive to the choice of other constants and in all settings we have considered it had numerical values in the range $c_{11}\sim0.19-0.53$.

Note that for small $\epsilon$ the exponent $A_3\sim\exp(-c_{11}/\epsilon)$ takes extremely small values. For static space-time the function $A$ defines the redshift effect $\lambda\sim A^{1/2}$, where $\lambda$ is the wavelength \cite{blau}.
Further we will term the exponential decay of $A$ as {\it red supershift}.

After the point $x_3$ the constant term in (\ref{eq_dadx}),(\ref{eq_dbdx}) dominates:
\begin{eqnarray}
&&-da/dx=db/dx=1,\label{eq_z3}
\end{eqnarray}
the solution has a form
\begin{eqnarray}
&&a=-x+c_{12},\ b=x+c_{13}.
\end{eqnarray}
Table \ref{tab_ranges} summarizes the investigated regimes and transitions.

\begin{table}\label{tab_ranges}
\begin{center}
\caption{Regimes, ranges and corresponding leading terms in the equations (\ref{eq_dadx}),(\ref{eq_dbdx}) for strong field, (\ref{eq_lin}) for weak field.}

\vspace{3mm}
\def\arraystretch{1.1}
\begin{tabular}{|c|c|c|}
\hline
regime&range&leading term\\ \hline
weak field, DM dominated&$x_{1a}<x<x_{uv2}$&$b\sim c_7$\\
weak field, Newton dominated&$x_{1b}<x<x_{1a}$&$b$\\
Schwarzschild&$x_{2}<x<x_{1b}$&$e^b$\\
supershift&$x_{3}<x<x_{2}$&$c_4e^{b-a}$\\
recovery from supershift&$x_{uv1}<x<x_{3}$&$1$\\
\hline
\end{tabular}

\end{center}
\end{table}

\paragraph*{UV regions.} Formally, the regime (\ref{eq_z3}) is preserved till $x\to-\infty$, leading to exponentially large values of $A$ and blue supershift. However, in physical scenarios the values of $r$ when this happens are exponentially small. Looking at $a$-profile, which near $r_s$ rapidly falls down till $a_3\sim-c_{11}/\epsilon$ then recovers at a rate linear in $x$, we can estimate a point where $a$ again approaches zero, as $r_{uv1}\sim r_s\exp(-c_{11}/\epsilon)$. For realistic $\epsilon$, this value is very small, in particular, it is much less then Planck length. We tend to view this innermost UV region as an artifact of the model, where the solution goes beyond its physical applicability. Quantum corrections or modifications, introducing an inner core in the matter distribution, can remove the innermost UV region from the model. Later we will consider several possibilities for such modifications.

A similar effect, an outermost UV region can be found, continuing the solution beyond $x_1$ point. At large $r$ the logarithmic term in potential is unbounded, so one will at first loose applicability of the weak field approximation and further, in strong field theory, find the values $A\gg1$. In physical scenarios this happens at exponentially large $r_{uv2}\sim r_s\exp(\alpha_{uv}/(2\epsilon))$, where the constant $\alpha_{uv}$ determines the strength of UV shift, e.g., $\alpha_{uv}=0.1$ corresponds to $\Delta A/A=0.1$ and $\Delta \lambda/\lambda=0.05$.
Again, substitution of physical $\epsilon$ gives the values of $r$ beyond the physical range, actually much larger than the size of the universe. In reality other effects will take place, e.g., reconnections of dark matter flows between different galaxies. Since these effects go beyond spherically symmetric approximation, they will definitely require an update of the model. 

\paragraph*{Naked singularity and cosmic censorship.} The behavior of $A,B$-functions in innermost UV region resembles that of Schwarzschild solution in the vicinity of so called naked singularity. It is commonly believed that in physically consistent models all singularities must be covered by event horizons, although this statement, called cosmic censorship principle, has never been strictly proven. The type of singularity, not covered by event horizon, can be created by setting in Schwarzschild solution $r_s$ to a negative value, formally corresponding to the case of negative mass. Although there is no event horizon in RDM solution either, the difference comes evident in detailed comparison of the profiles. Schwarzschild's $A\sim|r_s|/r$ at $r\to0$ produces strong UV factor, e.g., $A\sim10$ at $r\sim|r_s|/10$. For RDM solution the profiles at first suffer red supershift, making a recovery of solution to UV region possible only at $r\sim r_{uv1}$ and achieving $A\sim10$ at even smaller values $r\sim r_{uv1}/10$. For RDM solution this region becomes physically inaccessible. Due to the red supershift effect and removed UV region, RDM solution will look for an external observer completely black, just like a black hole. This explanation can serve as a relaxed version of cosmic censorship principle, applicable to RDM model.

\paragraph*{Dependence of solution on parameters.} RDM solution on Fig.\ref{f2},\ref{f4} is given for $\epsilon=0.04$, $c_2=1$, $c_3=1$. This setting makes a graphical representation of the solution convenient, while much smaller values of $\epsilon$ are of practical relevance. For smaller $\epsilon$ the overall shape of the solution remains the same, but all dependencies become sharper. In particular, the increase of $b$ near $r_s$ becomes larger and RDM solution on the right of $r_s$ as a whole tends to Schwarzschild solution. On the other hand, with decreasing $\epsilon$ the supershift effect becomes stronger and $ab$-profiles fall deeper and deeper in the negative region. Thus, RDM solution on the left of $r_s$ does not tend to Schwarzschild solution, no matter how small $\epsilon$ is.

Considering the effect of $c_{2,3}$, it is convenient to unify them in one constant $c_5=c_3/c_2^2$. Tachyonic case (TRDM) corresponds to $c_5>0$, null case (NRDM) is $c_5=0$ and massive case (MRDM) is $c_5<0$. In the last case practically relevant configurations start from faraway point $x_1$ with $A_1=1$, following a restriction $c_5>-1$. Large $c_2$ for TRDM and MRDM lead to $c_5\to0$ NRDM. In this transition the solutions are insignificantly deformed and all look like Fig.\ref{f2},\ref{f4}. The reason for such indifference of the result to the type of matter is that in the range of interest the function $A$ becomes very small and the system (\ref{eq_dAdx}), (\ref{eq_dBdx}) becomes independent on $c_5$, provided that the values of $c_5$ are bounded.

MRDM solutions have a peculiarity, related with a square root in the equations. When $A$ increases, the solution breaks in the point $A_{max}=-1/c_5$ and cannot be continued further. If $c_5$ is not too close to its lower limit $-1$, say $c_5>-0.9$, such a point belongs to the innermost or outermost UV region, i.e., normally is located beyond the model. If $c_5\to-1$, the breaking point approaches $x_1$ from outside. If $\epsilon$ in this transition is kept constant, the model exhibits increasing supershift, similar to small $\epsilon$ values. 

For TRDM, approaching $c_5\to+\infty$ at fixed $\epsilon$ also leads to an increasing supershift. In this limit the world lines on Fig.\ref{f1} right become horizontal, $t=Const$, representing the tachyonic flows with zero energy and non-zero momentum. This configuration was called in \cite{tachyo-dm} ``a tachyonic monopole''. Due to $T$-symmetry and stationarity of the problem, the world lines in this configuration remain horizontal also in strong fields. The equations in this limit have a form:
\begin{eqnarray}
&&da/dx=-1+e^b+2\epsilon e^{b-a/2},\ db/dx=1-e^b,\label{eq_trdm0}
\end{eqnarray}
their solution can be found analytically:
\begin{eqnarray}
&&b=-\log \left(1-{r_s} e^{-x}\right),\ d=\sqrt{e^x-{r_s}},\\
&&a=2 \log \left(\frac{2 d {\epsilon} \sqrt{{r_1}}}{\sqrt{{r_1}-{r_s}}}-2 d
   {\epsilon} \log \left(\sqrt{{r_1}-{r_s}}+\sqrt{{r_1}}\right)\right.\\
&&\left.+2 d
   {\epsilon} \log \left(d+e^{x/2}\right)+\frac{d
   \sqrt{{r_1}}}{\sqrt{{r_1}-{r_s}}}-2 {\epsilon} e^{x/2}\right)-x.
\ \nn
\end{eqnarray}
Here $b$ coincides with that of Schwarzschild solution, since $b$-equation in (\ref{eq_trdm0}) does not have DM-term. 
The solution for $a$ is defined at $r>r_2>r_s$. The limit $r_2$ becomes closer to $r_s$, as one considers fixed $r_s,r_1$, while $\epsilon\to0$. The solution cannot be continued to smaller $r$. Approaching this solution along $c_5\to+\infty$ path explains the reason: there is a supershift, which makes $ab$-profiles falling down to minus infinity after $r_2$.

\paragraph*{Details on the numerical procedure.} The system (\ref{eq_dadx}),(\ref{eq_dbdx}) is autonomous. At first, this means that one solution can be used to generate other solutions by the translation $x\to x+c$, equivalent to scaling of $r$-coordinate. Secondly, one can transform the system to a single differential equation
\begin{equation}
db/da=(1-e^b+c_4e^{b-a}/\sqrt{1 + c_5e^a})/(-1+e^b+c_4e^{b-a}\sqrt{1 + c_5e^a}),\label{eq_dbda}
\end{equation}
taking either $a$ or $b$ as an independent variable. After solving this equation, the profile of $x$ can be then found by an integration of (\ref{eq_dadx}) or (\ref{eq_dbdx}). Since both $a$- and $b$-profiles appear to be non-monotonous, it is convenient to separate the integration of (\ref{eq_dbda}) to two parts. From point $x_1$ to an intermediate point $x_{2a}$, located between $x_2$ and $x_3$, take $a$ as independent variable and solve the equation (\ref{eq_dbda}) for $b(a)$. Then, after $x_{2a}$, take $b$ as independent variable and solve the equation (\ref{eq_dbda}) for $a(b)$. The equation in form (\ref{eq_dbda}) is most convenient for the numerical solution, since large exponents enter both in nominator and denominator of (\ref{eq_dbda}), leading to their cancellation. As a result, most of the time the r.h.s. of the equation equals either $1$ or $-1$, while the solution follows a regime with one dominating term, $b=Const\pm a$. The regime change creates a rapid variation of the r.h.s. between $1$ and $-1$. Therefore, the equation belongs to stiff type, requiring special measures for its integration.
We use Mathematica \cite{math}, algorithm {\it NDSolve}, combining an automatic detection of stiffness, a balanced switching between explicit and implicit methods of integration and an adaptive selection of integration step. To test the precision, we performed the same integration, enforcing 10 times smaller integration step, and obtain a numerical difference of solutions of the order $10^{-6}$.

\section{Structure of geodesics}\label{sec_geode}
Let us investigate the structure of geodesics in the obtained space-time geometry. The general geodesics are trajectories of probe particles, carrying a negligible energy-momentum, which does not contribute in the gravitational field and should not be taken into account in Einstein field equation. There are also exceptional geodesics, which coincide with those of radial dark matter and are already taken into account.

\begin{figure}
\centering
\includegraphics[width=0.49\textwidth]{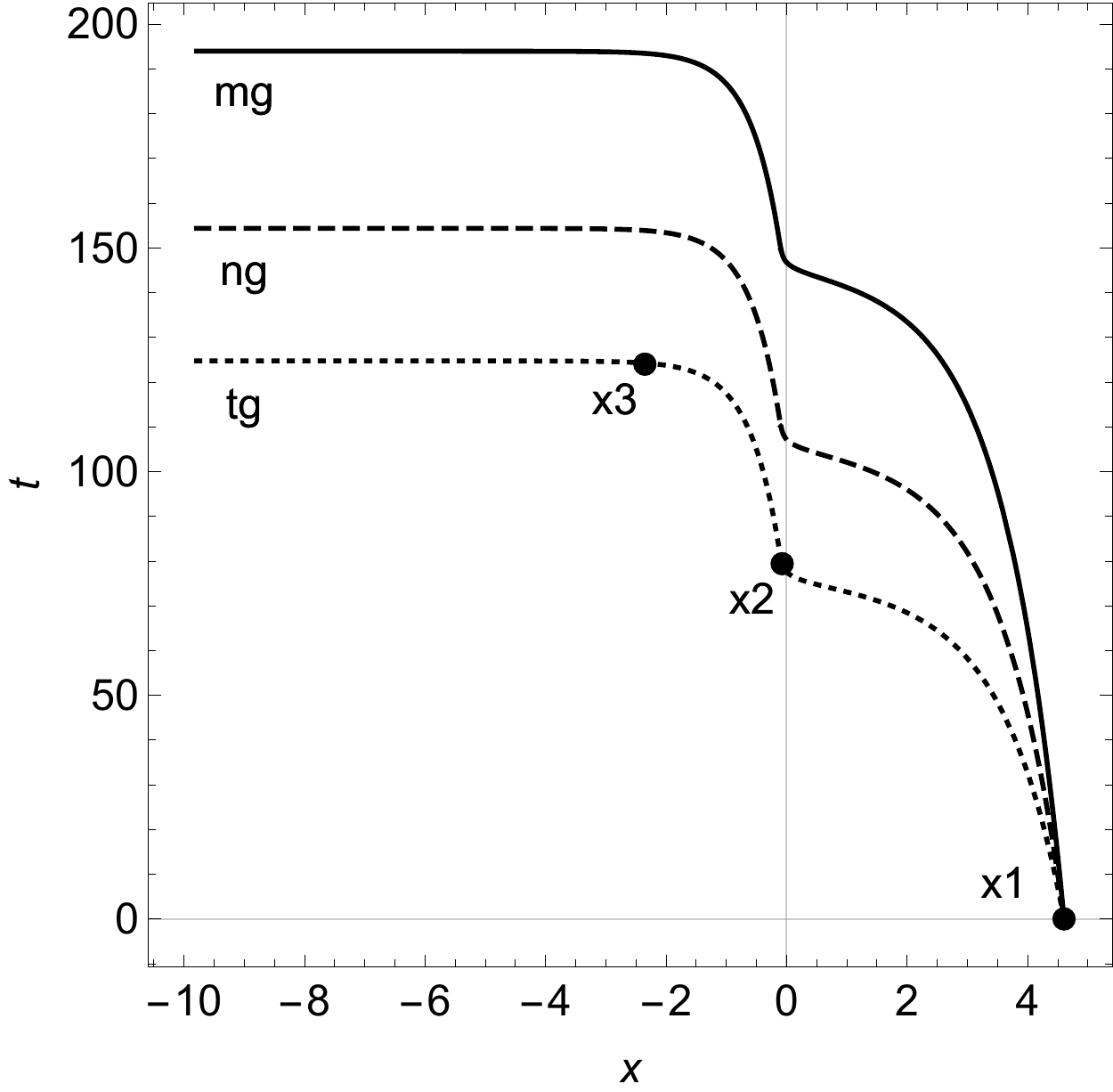}
\includegraphics[width=0.49\textwidth]{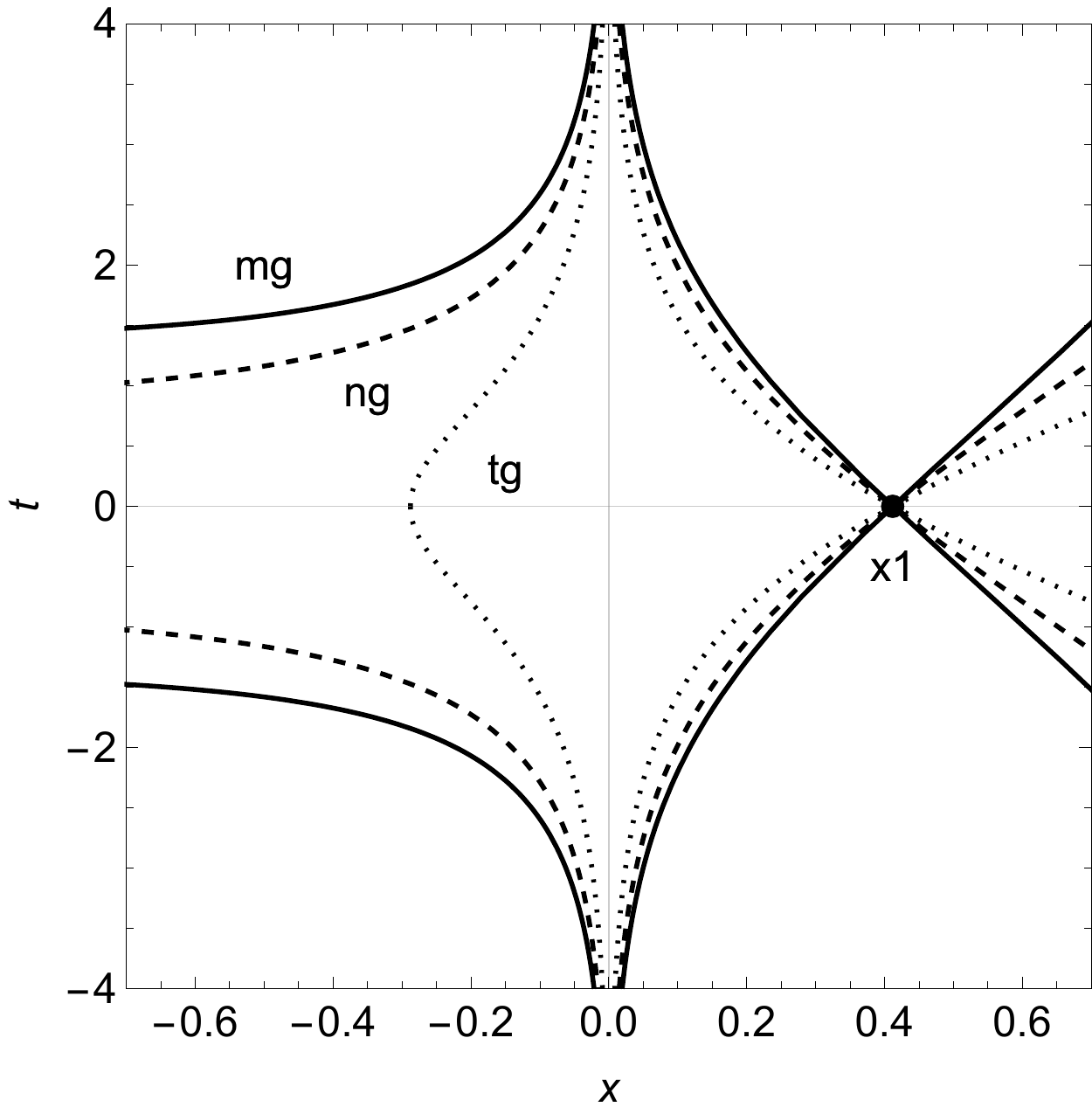}
\caption{Radial geodesics of tachyonic, null and massive type (tg, ng, mg). Vertical axis is time $t$, horizontal axis is $x=\log r$. On the left: RDM solution with $\epsilon=0.01$, on the right: Schwarzschild solution. The geodesics in Schwarzschild solution go to infinity, corresponding to time freeze at event horizon, then (ng,mg) fall into the singularity, while (tg) is repelled by it. Time-reflected geodesics correspond to the white hole solution. RDM geodesics have a finite delay time between the points $x_2$ and $x_3$ and then go towards the singularity.}
\label{f5}
\end{figure}

\paragraph*{Radial geodesics.} These geodesics are defined by equations (\ref{eq_geode}), with $c_1=0$ and own constants $\tilde c_2$, $\tilde c_3$. The shape of the geodesics is defined by the equation
\begin{eqnarray}
&&dt/dr=\pm u^t/u^r=\pm\sqrt{B/A}/\sqrt{1+\tilde c_5 A}
\end{eqnarray}
with $\tilde c_5=\tilde c_3/\tilde c_2^2$, or
\begin{eqnarray}
&&dt/dx=\pm e^{x+(b-a)/2}/\sqrt{1+\tilde c_5 e^a}.\label{eq_geode2}
\end{eqnarray}
The result of numerical integration of this equation is shown on Fig.\ref{f5} left. The gravitational field was computed for $\epsilon=0.01$, $c_5=1$ (TRDM). The geodesics are computed for
\begin{itemize}
\item $\tilde c_5=1$, tachyonic geodesics (tg), actually coincident with those of TRDM,
\item $\tilde c_5=0$, null geodesics (ng),
\item $\tilde c_5=-0.5$, massive geodesics (mg).
\end{itemize}
The whole structure of geodesics can be completed by continuous variation of $\tilde c_5$, $T$-reflections $t\to -t$ and shifts $t\to t+c$. 

The geodesics, shown on Fig.\ref{f5} left, have a common feature, a slowdown between the points $x_2$ and $x_3$. This slowdown is caused by the increased value of $b-a$ in the supershift region, leading to the increased derivative in (\ref{eq_geode2}). The effect is almost independent on the matter type, since the values of $a$ in this regime go to deep negative region. As a result, the equation (\ref{eq_geode2}) becomes insensitive to the constant $\tilde c_5$, provided that this constant remains bounded. The value of temporal shift $t_3-t_2$ increases with decreasing $\epsilon$, so we have specially reduced the value of $\epsilon$ for this plot, to present this effect better.

Considering geodesics parametrized by proper time (for mg) or proper length (for tg):
\begin{eqnarray}
&&d\tau/dr=\pm 1/u^r=\pm\sqrt{BA}/\sqrt{1+\tilde c_5 A}/c_2,\\
&&d\tau/dx=\pm e^{x+(b+a)/2}/\sqrt{1+\tilde c_5 e^a}/c_2,
\end{eqnarray}
we will see that after the point $x_2$ the function $b+a$ rapidly falls in the negative region, as a result, $\tau$ after this point will possess a negligible variation.

The delay $t_3-t_2$ is determined by the difference $b-a$ which though large, is still much less then the range of $a,b$ values in the supershift zone. Slowdown factor $\Delta t/\Delta\tau$ for a stationary observer, staying at constant $r$, is defined by the redshift function $a$, while the slowdown factor for freely falling observer is defined by $(b+a)/2$. These factors are very large, about $(A_3)^{-1/2}\sim\exp(c_{11}/(2\epsilon))$.

Fig.\ref{f5} right shows geodesics for Schwarzschild solution. At first, Schwarzschild geodesics go to $t\to+\infty$, showing well known effect of time freeze at the horizon, from the viewpoint of a distant observer. This is a coordinate singularity, which can be overcomed by transformation to a different time coordinate, e.g., proper time for one of the geodesics. Formally, the geodesics can be also continued in standard coordinates, as shown on Fig.\ref{f5} right. After passing the horizon they fall down from the infinity and reach the central singularity in a finite time, for the geodesics of massive and null type. The geodesics of tachyonic type behave differently, they repel from the central singularity and stop approaching it at some finite distance. Time reflected geodesics of all types go from the central region to minus infinity in time and then, after passing the horizon, go to the outer region. These geodesics correspond to the white hole solution, inherently present in $T$-symmetric Schwarzschild scenario. These solutions can be also considered in so called maximal topological extension of Schwarzschild scenario. One takes two copies of Fig.\ref{f5} right and connects the geodesics at $t\to+\infty$ on each copy and at $t\to-\infty$ in cross-like manner between the copies. After that the geodesics starting from the central singularity go to a different copy, ``a mirror universe''. There are also geodesics, coming from the copy of the white hole to our side. This solution, ``eternal black hole'' is commonly considered as an artifact or an idealized case, never or rarely encountered in the reality. In real collapse scenarios, the mirror universe and the white hole are cut off (``exorcised'' \cite{blau}) from the solution, leaving there only $T$-nonsymmetric black hole.

There are severe distinctions between RDM and Schwarzschild geodesics. For RDM standard $(t,r)$ coordinates are non-singular at $r>0$ and are applicable to the whole solution. There is one global patch in the solution, which does not require topological extensions and further cuts. The slowdown in RDM supershift region is a remnant of Schwarzschild time freeze, however for RDM it lasts finite time. After passing the supershift region, RDM geodesics rapidly go to central singularity. The difference is that for RDM null and tachyonic geodesics reach it in a finite time, while for Schwarzschild null and massive geodesics do so.

Massive geodesics in RDM model possess already known peculiarity. The geodesic with $\tilde c_5<0$ does not reach the central singularity, but at $A_{max}=-1/\tilde c_5$ has a turning point with a vertical tangent (\ref{eq_geode2}). Being connected to $T$-reflected solution in this point, the geodesics returns to the outer region and in the range $x>x_1$, not shown on Fig.\ref{f5} left, has a second turning point. Formally, the whole solution for massive radial geodesic is periodic. On the other hand, for the geodesics starting faraway from the center with $-0.9<\tilde c_5<0$, the both turning points are located in UV regions. According to the previous discussion, UV regions have to be removed from the model. As a result, the considered (mg) family behaves in the same way as (ng,tg), reaching the smallest and the largest distances available in the model in a finite time.

\begin{figure}
\centering
\includegraphics[width=0.49\textwidth]{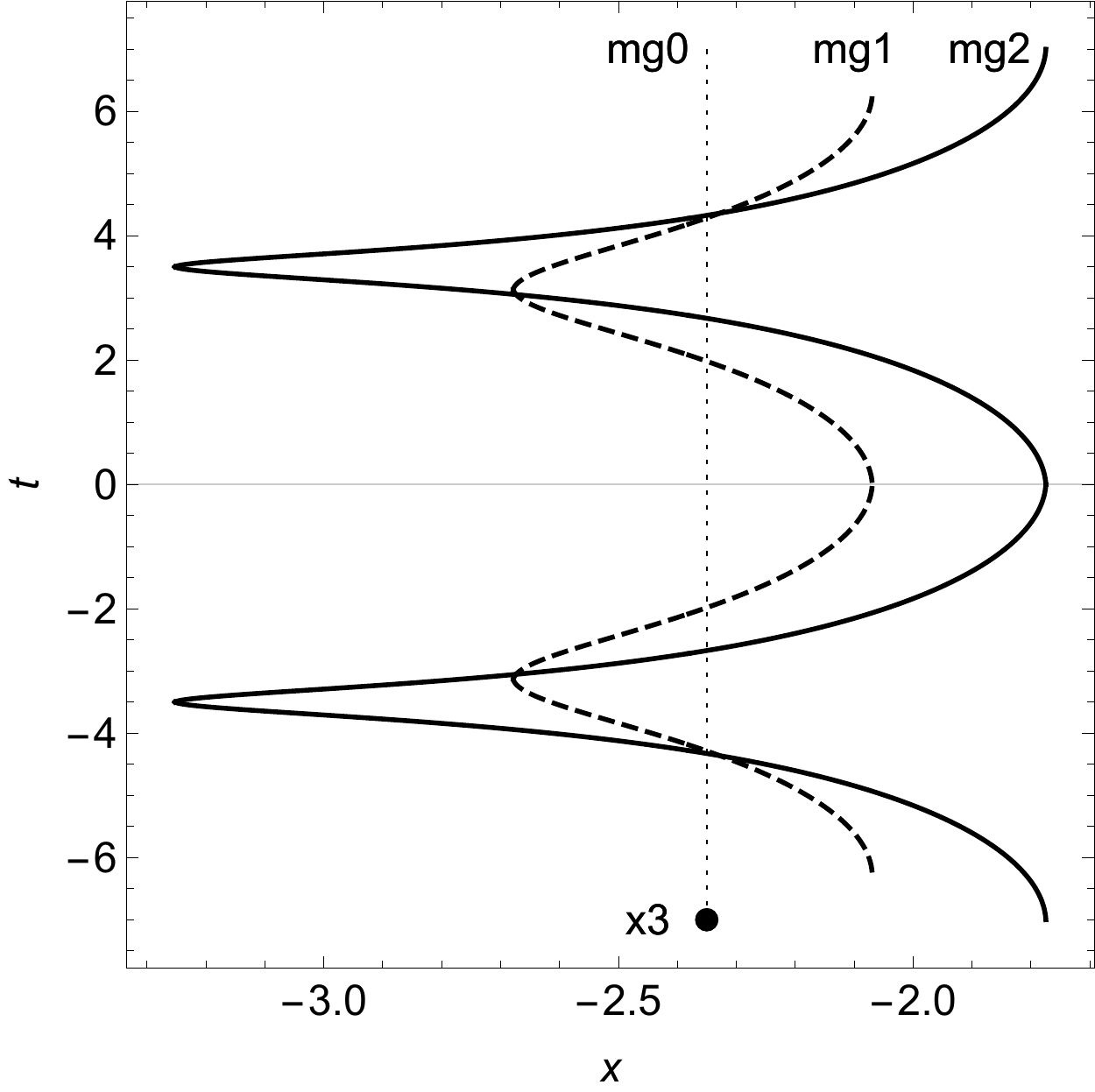}
\includegraphics[width=0.49\textwidth]{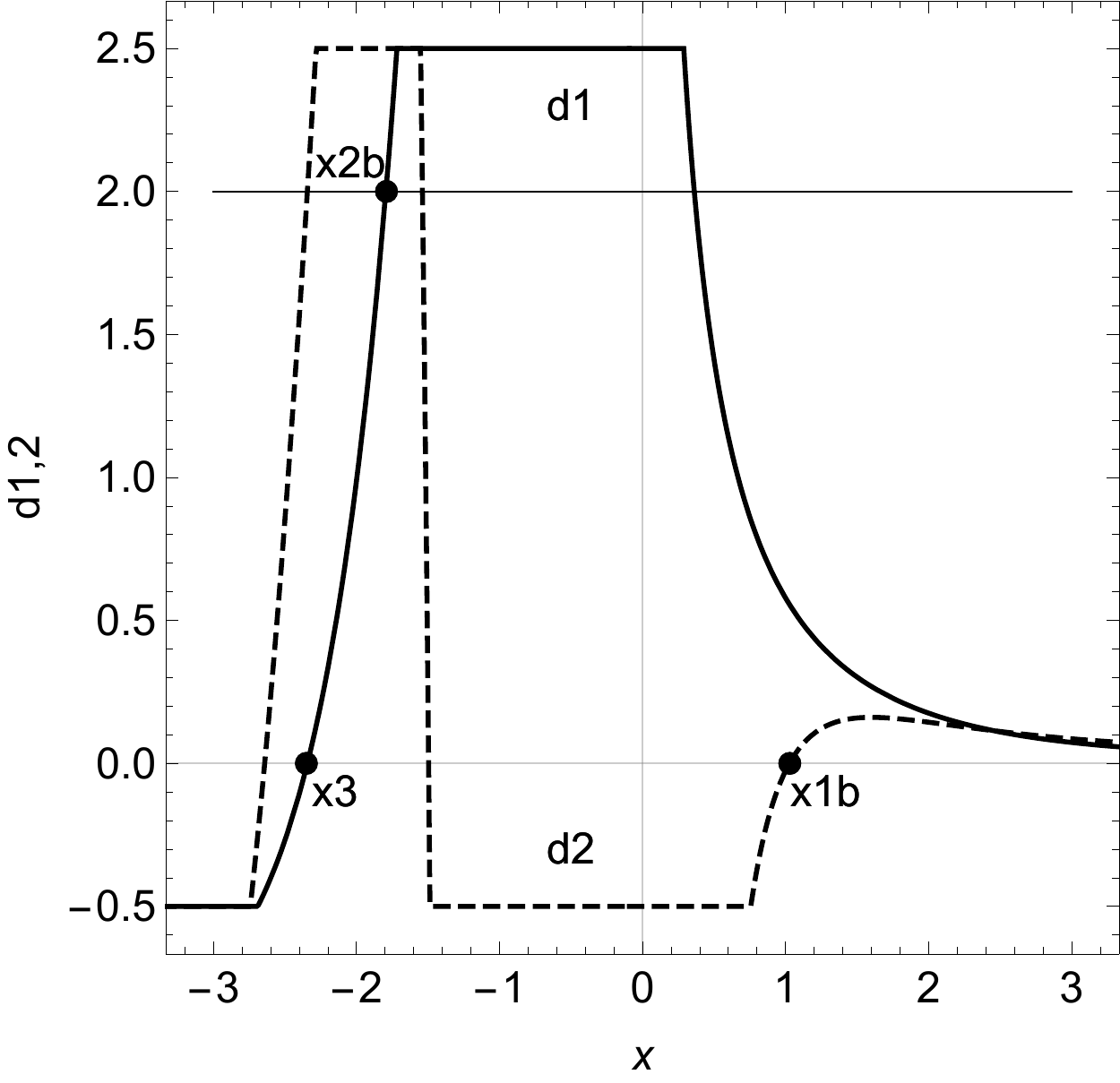}
\caption{On the left: periodic geodesics of massive type in RDM solution. Vertical axis is time $t$, horizontal axis is $x=\log r$. On the right: evolution of $d_{1,2}$ functions, defining stability of circular orbits. The regions $x>x_{1b}$ and $x_3<x<x_{2b}$ correspond to stable circular orbits.}
\label{f6}
\end{figure}

The periodic (mg) solutions are available in the vicinity of $x_3$ point, where the function $a$ has a minimum. Such solutions are shown on Fig.\ref{f6} left. The point $x_3$ itself corresponds to a static equilibrium solution (mg0) with $\tilde c_5=-1/A_3$, while the other oscillating solutions (mg1,mg2) have larger $\tilde c_5$.

\paragraph*{Circular geodesics.} Following \cite{0607125}, introduce a Lagrangian for the probe particle in a background gravitational field:
\begin{eqnarray}
&&L=\half(-e^a\dot t^2+e^b\dot r^2+r^2\dot\theta^2+r^2\sin^2\theta\,\dot\phi^2),
\end{eqnarray}
where dot denotes the derivative with respect to proper time. Further restrict the trajectories to the equator $\theta=\pi/2$.
The generalized momenta have a form:
\begin{eqnarray}
&&p_t=-E=-e^a\dot t,\ p_r=e^b\dot r,\ p_\phi=W=r^2\dot\phi,
\end{eqnarray}
the symmetries of the Lagrangian imply the conservation of $W,E$. The evolution of the radial coordinate is described by the equation:
\begin{eqnarray}
&&\dot r^2+V_{\mbox{\small eff}}(r)=0,\ 
V_{\mbox{\small eff}}(r)=e^{-b(r)}\left(1+\frac{W^2}{r^2}-E^2e^{-a(r)}\right),
\end{eqnarray}
while stable circular orbits are defined by
\begin{eqnarray}
&&V_{\mbox{\small eff}}=0,\ \frac{dV_{\mbox{\small eff}}}{dr}=0,\ \frac{d^2V_{\mbox{\small eff}}}{dr^2}>0,
\end{eqnarray}
or equivalently
\begin{eqnarray}
&&V_{\mbox{\small eff}}(x)=e^{-b(x)}\left(1+W^2e^{-2x}-E^2e^{-a(x)}\right),\\
&&V_{\mbox{\small eff}}=0,\ \frac{dV_{\mbox{\small eff}}}{dx}=0,\ \frac{d^2V_{\mbox{\small eff}}}{dx^2}>0.
\end{eqnarray}
Further in this section prime denotes the derivative with respect to $x$. Solving $V_{\mbox{\small eff}}=0$, $V'_{\mbox{\small eff}}=0$ at fixed $x$ for $W^2$,$E^2$:
\begin{eqnarray}
&&W^2=e^{2x}a'/(2 - a'),\ E^2=2e^a/(2 - a'),
\end{eqnarray}
and substituting into
\begin{eqnarray}
&&V''_{\mbox{\small eff}}=2e^{-b}(a'(2-a')+a'')/(2 - a'),
\end{eqnarray}
introduce two functions:
\begin{eqnarray}
&&d_1=a',\ d_2=a'(2-a')+a''
\end{eqnarray}
which define stability of circular orbits by the following inequalities:
\begin{eqnarray}
&&0\leq d_1\leq2,\ d_2\geq0.
\end{eqnarray}
Further, according to \cite{0607125}, the velocity of the particle on the circular orbit has a form:
\begin{eqnarray}
&&v^2=a'/2,
\end{eqnarray}
so that $0\leq d_1\leq2$ becomes equivalent to $0\leq v\leq1$. Let us consider these expressions for the following limits:

\begin{itemize}
\item Newtonian limit, all orbits are stable:
\begin{equation}
a=-r_se^{-x},\ d_1\sim d_2\sim r_s/r\ll1,\ v^2=r_s/(2r).
\end{equation}
\item Schwarzschild limit, the orbits above ISCO $r > 3 r_s$ are stable:
\begin{equation}
a=\log(1-r_se^{-x}),\ d_1=r_s/(r - r_s),\ d_2=(r - 3 r_s) r_s/(r - r_s)^2.
\end{equation}
\item Weak field DM-dominated limit, all orbits are stable:
\begin{equation}
a=Const+2\epsilon x,\ \epsilon\ll1,\ d_1=2\epsilon,\ d_2\sim4\epsilon,\ v^2=\epsilon.
\end{equation}
\end{itemize}

Generally, the situation for RDM is shown on Fig.\ref{f6} right. The profiles $d_{1,2}$ are clamped by a function $\mbox{clamp}(d)=\max(\min(d,2.5),-0.5)$, to show their behavior in the region of interest. Since the function $a$ has the minimum in the point $x_3$, in its right vicinity $a''>0$ and $a'\to+0$. As a result, the second island of stable circular orbits appears (thereby ISCO is not the innermost one anymore). The velocity $v$ in this island follows the profile of $d_1$ and  in the range between $x_3$ and $x_{2b}$ varies from $0$ to $1$, the speed of light.

\section{Matching the model with a real galaxy}\label{sec_galaxy}

\begin{figure}
\centering
\includegraphics[width=0.8\textwidth]{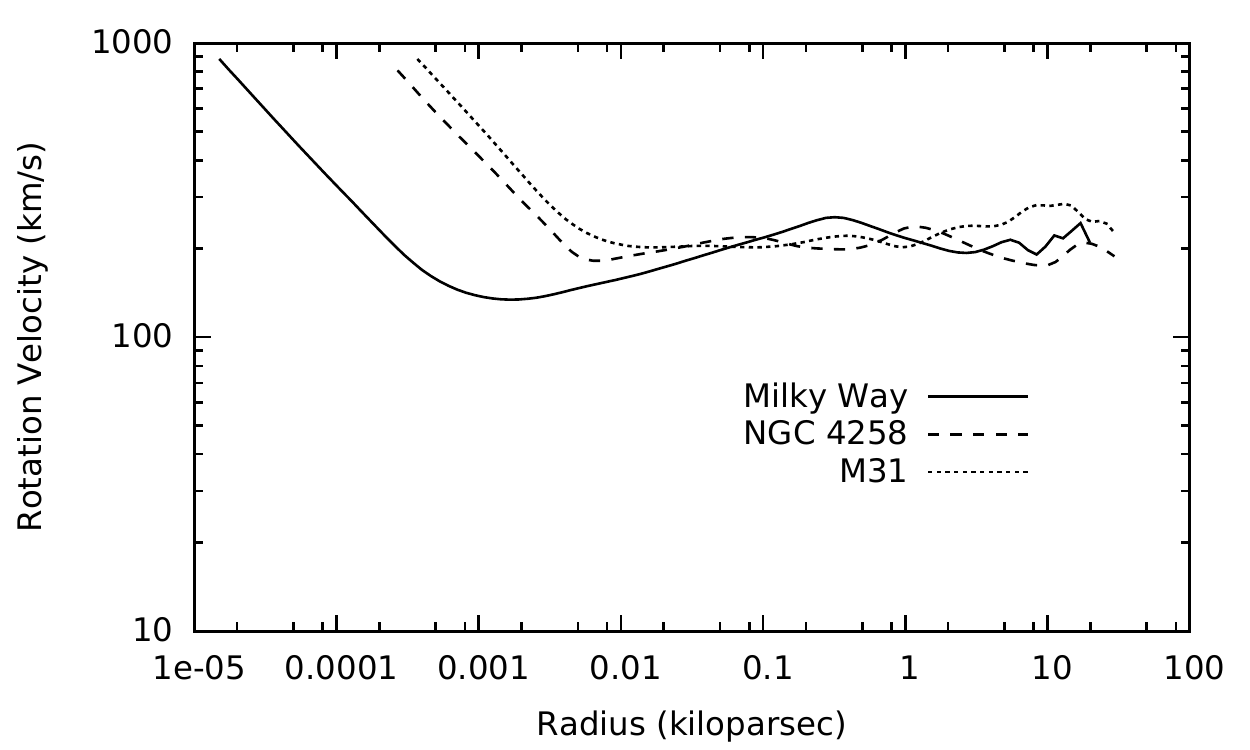}

\caption{The measured rotation curves for three galaxies, data from \cite{VR2}.}
\label{f7}
\end{figure}

Fig.\ref{f7} shows experimental rotation curves for three galaxies \cite{VR2}. In asymptotic region these curves go to approximately the same constant and in RDM model they possess the same value $\epsilon\sim4\cdot10^{-7}$. For our galaxy, Milky Way, the estimations for the central black hole are also available \cite{ghez}. This allows to reconstruct all ranges of RDM model, shown in Table \ref{tab_mw}. The reconstruction is done for $c_5=1$ TRDM, while other settings of this parameter give similar numbers, provided that it is not selected too close to its limiting values. The full list of model constants then reads
\begin{eqnarray}
&&c_1=9.42809\cdot10^{-8},\ c_2=c_3=1,\ c_4=3.77124\cdot10^{-7},\\ 
&&c_5=1,\ c_6=5.33333\cdot10^{-7},\ c_7=2.66667\cdot10^{-7}. 
\end{eqnarray}

\begin{table}\label{tab_mw}
\begin{center}
\caption{Model parameters and ranges for the Milky Way galaxy}

\def\arraystretch{1.1}
\begin{tabular}{|c|c|}
\hline
model parameters&$\epsilon=4\cdot10^{-7}$, $r_s=1.2\cdot10^{10}$m\\ \hline
a border of the galaxy&$r_1=3.1\cdot10^{21}$m,\\ 
(starting point)& $a_1=0$, $b_1=2.67\cdot10^{-7}$\\ \hline
data at Earth location&$r_E=2.57\cdot10^{20}$m, \\ 
&$a_E=-2\cdot10^{-6}$, $b_E=b_1+4\cdot10^{-11}$\\ \hline
switch from DM-dominated&$r_{1a}=1.5\cdot10^{16}$m, \\ 
to Newtonian regime&$a_{1a}=-1.05\cdot10^{-5}$, $b_{1a}=9.91\cdot10^{-7}$\\ \hline
switch to Schwarzschild regime,&$r_{1b}=3.33\cdot10^{10}$m, \\ 
end of outer stable circular orbits&$a_{1b}=-b_{1b}-2.06\cdot10^{-5}$, $b_{1b}=0.404$\\ \hline
begin of the supershift&$r_{2}=1.11\cdot10^{10}$m, \\ 
&$a_{2}=-14.79$, $b_{2}=13.40$\\ \hline
(switch of integration $b(a)\to a(b)$)&$r_{2a}=r_2-1.2\cdot10^{4}$m, \\ 
&$a_{2a}=-16.79$, $b_{2a}=12.54$\\ \hline
begin of inner stable circular orbits&$r_{2b}=1.2\cdot10^7$m, \\ 
&$a_{2b}=b_3-14.34$, $b_{2b}=b_3+1.55$\\ \hline
end of inner stable circular orbits,&$r_{3}=6.8\cdot10^6$m, \\
end of the supershift&$a_{3}=b_3-14.79$, $b_{3}=-1.33\cdot10^6$\\ \hline
variation of redshift at the & $r_{Pl}=1.62\cdot10^{-35}$m, \\ 
minimal radius (Planck length)&$a_{Pl}/a_{3}-1=-7.19\cdot10^{-5}$\\ \hline
estimated UV-shift at the & $r_{M31}=2.44\cdot10^{22}$m,  \\ 
maximal radius (Andromeda galaxy)& $a_{M31}=1.65\cdot10^{-6}$\\ \hline
temporal shift&$t_3-t_2=1.55$ years\\ \hline
\end{tabular}

\end{center}
\end{table}

The integration starts at the outer limit of the galaxy, selected at $r_1=100$kpc from the center. The rotation curve of the Milky Way is measured till this and even larger distances \cite{13102659}. We set here $a_1=0$, meaning that global time is measured by the clock of an observer, located at this distance. The clock at Earth location $r_E=8.3$kpc deviate from this clock by a negligible correction factor $a_E=-2\cdot10^{-6}$. Initially the gravitational field is DM-dominated, then it switches to Newtonian dominated at $r_{1a}=0.5$pc. At this point, approximately, the curve on Fig.\ref{f7} changes its behavior.
Further, at $r_{1b}=3.33\cdot10^{7}$km the field switches to Schwarzschild dominated and stable circular orbits end here. The closest stars to the central black hole, so called S-stars, from which the best measured S0-2 and S0-102 move on strongly elliptic orbits in the range between $r\sim2\cdot10^{13}$km and $r\sim3\cdot10^{14}$km. This range is located well below $r_{1a}$ and well above $r_{1b}$. This means that corrections both from dark matter and general relativity in this range are small and the orbits of S0-2 and S0-102 stars are approximately Keplerian, in agreement with current observations \cite{Gillessen,Meyer}.
At $r_2=1.11\cdot10^{7}$km, approximately 16 solar radii, the function $b$ reaches a maximum and typical for RDM model supershift regime begins. The position of $b(a)$ maximum is located close to theoretical estimation $a_2=\log c_4$. The value $r_2$ is located a bit below the nominal value $r_s$. However, the other features of the gravitational field in this region are related to $r_2$, e.g., $r_{1b}=3r_2$, meaning that in the range of Schwarzschild domination $r_2$ plays a role of corrected gravitational radius. Further, at $r_{2b}=1.2\cdot10^4$km stable circular orbits appear again. At $r_{3}=6.8\cdot10^3$km, approximately Earth radius, stable circular orbits disappear and supershift ends. The $ab$-values reached there are about $a_{3}\sim b_{3}=-1.33\cdot10^6$, close to theoretical estimation $b_3=-0.5/c_4$. In further decrease of the radius $ab$-values remain almost constant in relation to this value, till the lower limit at Planck length. At the upper limit, for which we select the distance to the next major galaxy $r_{M31}=790$kpc, $a$-value increases only slightly, providing a negligible UV-shift $a_{M31}=1.65\cdot10^{-6}$. The geodesics have a temporal shift $t_3-t_2=1.55$ years, while crossing the supershift region. The coordinate length of the supershift region is about $37$ light seconds, comprising a slowdown factor of $1.32\cdot10^6$.

\vspace{3mm}
The experimental rotation curves possess other features, peaks and dips, deviating it from the idealized flat line. These features require more sophisticated modeling, involving the distribution of the luminous matter. Also, the central black hole is probably not the single source of dark matter in the galaxy and the contribution of other black holes should be also taken into account. To complete this paper, let us consider the last possibility in more details.

\begin{figure}
\centering
\includegraphics[width=0.8\textwidth]{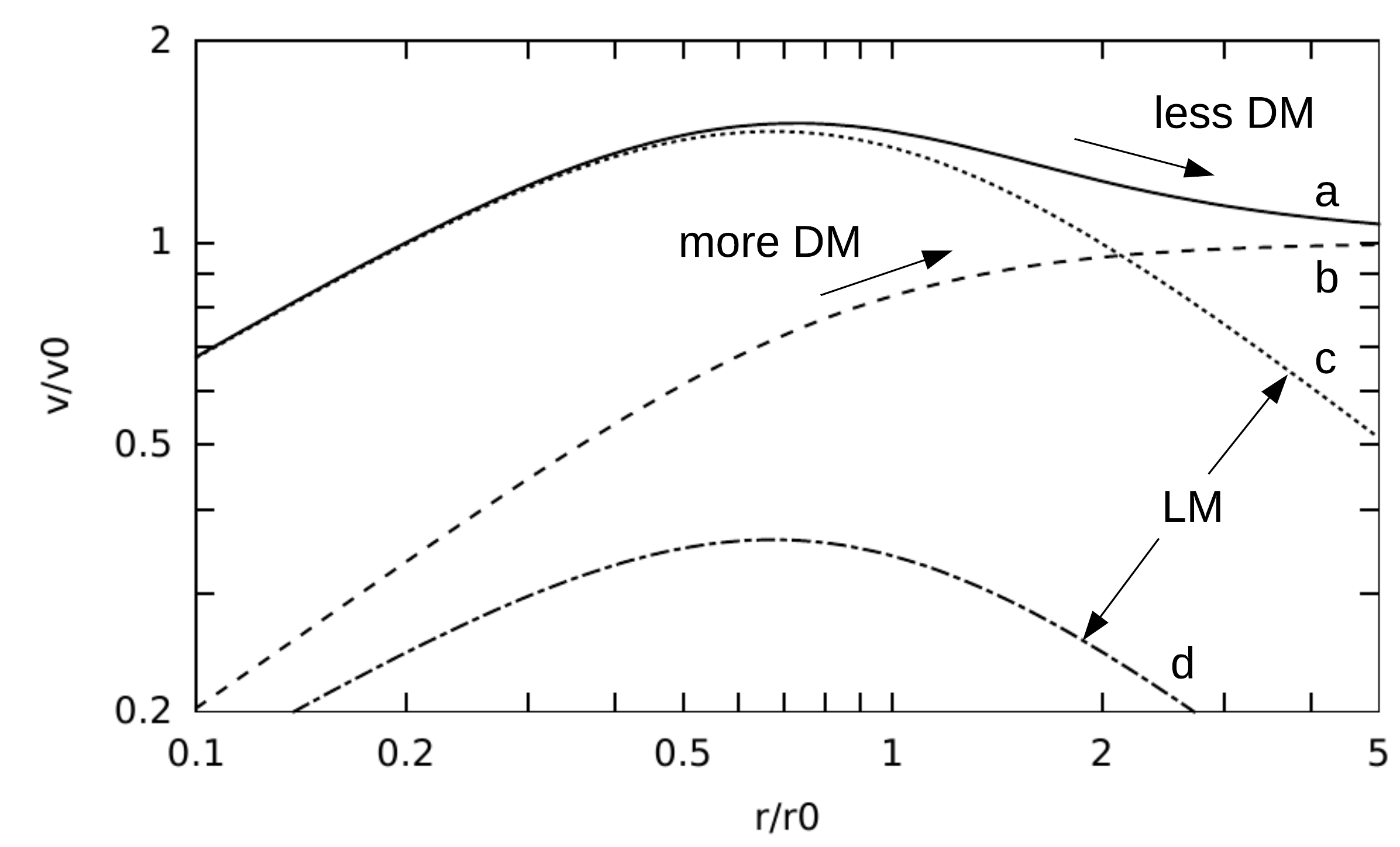}

\caption{Universal rotation curves (URC), data from \cite{salucci}. On axes: radius and velocity, in relative units. The curve (a) shows URC for large relative luminosity ($L/L_0=3$), corresponding to the galaxies with a smaller content of dark matter. The curve (b) is URC for small relative luminosity ($L/L_0=0.06$), corresponding to the galaxies with a greater content of dark matter. The curves (c) and (d) show the contribution of luminous matter (LM).}
\label{f8}
\end{figure}

\paragraph*{RDM model with distributed sources.}
The paper \cite{salucci} presents universal rotation curves (URC) of spiral galaxies, obtained by averaging of $\sim$1100 experimental velocity profiles, as functions of the distance to the center and the luminosity of the galaxy. Fig.\ref{f8} shows typical URC in relative units $r_0$ and $v_0$ (defined individually for every galaxy). The difference in the behavior of the curves on Fig.\ref{f7} and Fig.\ref{f8} is related to the fact that Fig.\ref{f7} covers a larger range of $r$ and includes the region of influence of the central black hole, while Fig.\ref{f8} does not cover this region.  For large luminosities (case a, $L/L_0=3$) URC passes through the maximum, then falls down and tends to the asymptotic value from above. For small luminosities (case b, $L/L_0=0.06$) URC monotonously increases and tends to the asymptotic value from below. The curves (c,d), taking into account only luminous matter, are located below URC and rapidly fall at large $r$.

Such behavior of rotation curves can be explained by the presence of the other black holes, which emit and absorb dark matter flows and are distributed over the galaxy. In this case the density of dark matter sources depends on the distance to the center of the galaxy. In weak field approximation the sources contribute additively to the gravitational potential. Considering a sum of independent isotropic sources, emitting dark matter flows to infinity, distributed in the galactic disk isotropically over the angle, we can show that the dark matter contribution will lead to increasing rotation curves. Indeed, the potential $\varphi\sim\log \,r$ is coincidentally a fundamental solution of Poisson equation in two dimensions, similar to $\varphi\sim r^{-1}$, a  fundamental solution of Poisson equation in three dimensions. As a result, if the sources of dark matter are distributed radially in the galactic plane as $\epsilon(r)>0$, then the gravitational acceleration in the galactic plane is defined by a cumulative value 
\begin{eqnarray}
&&{\cal E}(R)={\cal E}_{0}+2\pi\int_0^R dr \,r\,\epsilon(r)
\end{eqnarray}
and equals 
\begin{eqnarray}
&&a_{r}(R)={\cal E}(R)/R,
\end{eqnarray}
similar to spherically symmetric mass distribution in three dimensions, where the gravitational acceleration is defined by cumulative mass 
\begin{eqnarray}
&&M(R)=M_{0}+4\pi\int_0^R dr \,r^2\rho(r) 
\end{eqnarray}
and equals 
\begin{eqnarray}
&&a_{r}(R)=G_NM(R)/R^2.
\end{eqnarray}
Positive constants ${\cal E}_{0},M_{0}$ take into account possible contributions of the central singularity. Note that for radial planar mass distribution $\varphi\sim r^{-1}$ is not the fundamental solution of Poisson equation in two dimensions, the gravitational acceleration is not defined by cumulative values and is described by more complicated formulae \cite{salucci}. 

Therefore, for the square of orbital velocity we have
\begin{eqnarray}
&&v^{2}(R)={\cal E}(R)+v_{n}^{2}(R),
\end{eqnarray}
where the contribution of dark matter is described by an increasing function ${\cal E}(R)$. The function $v_{n}^{2}(R)$ includes luminous matter and Newtonian mass contributions from the central supermassive black hole and other black objects in the galaxy. The contribution of luminous matter has a typical maximum, shown on Fig.\ref{f8}. For highly luminous galaxies the contribution of luminous matter prevails in the vicinity of the maximum, as a result, the total rotation curve has a maximum. For low luminous galaxies the dark matter contribution prevails and the rotation curve monotonously increases.

We see that qualitative behavior of rotation curves of spiral galaxies can be described by RDM model with distributed sources. For more detailed modeling one requires the data or reasonable model assumptions on the spatial distribution of black holes in the galaxies. One will also need the relation of dark matter flow densities with parameters of the black holes, e.g., in the form of $r_s(\epsilon)$ dependence.

\paragraph*{Other extensions of RDM model.} The model possesses two ultraviolet cut limits, defining the points where (or better much earlier) it should be connected to the other models. It is known that at large scale the dark matter forms network-like superstructures, consisting of filaments that connect the galaxies. These structures can be composed of the dark matter world lines, stretched between galactic black holes. The superstructures can be considered in the next level model, which connects to the local models of the galaxies on their upper limit. This limit is located faraway from gravitational centers and one can consider the space-time as flat in the first approximation. A model of this kind has been considered in \cite{cosmic-web}. It constructs randomized networks in Minkowski space-time, composed of the world lines of relativistic particles, with the action equal to a linear combination of Minkowski length of the world lines. At appropriate choice of control parameters, it allows efficient computation of the networks of high complexity. The computational performance depends in a critical way on the presence of tachyons in the network, the model including both tachyonic and normal matter appears to be linear and more simple for computations.

\begin{figure}
\centering
\includegraphics[width=0.6\textwidth]{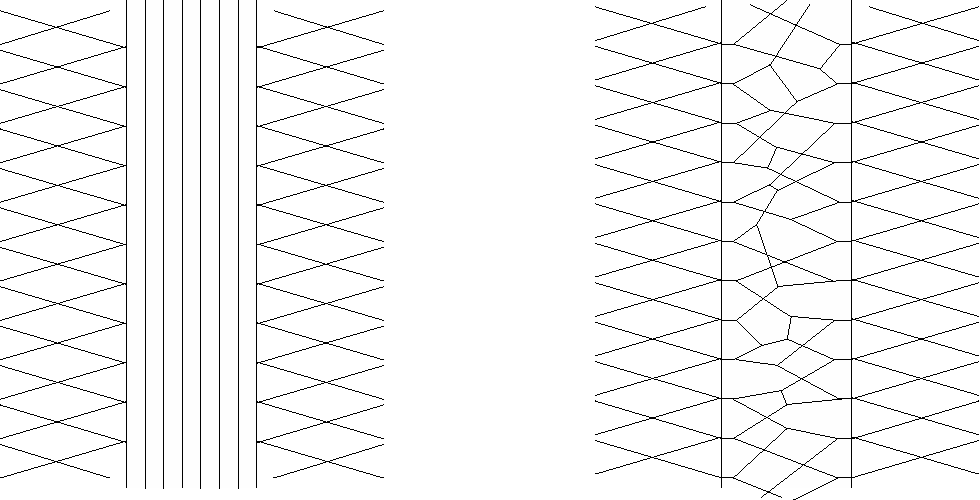}
\caption{RDM model extended by the inner core. On the left: inner core composed of static perfect fluid, on the right: inner core comprising a randomized network of interacting relativistic particles.}
\label{f9}
\end{figure}

At the lower limit, presumably below $r_2$ and well above $r_{Pl}$, the model can be modified by prescribing a different matter distribution or a different topology of the space-time. This approach allows to use the framework of classical mechanics and standard general relativity, not requiring the fundamental changes. The following extensions can be considered.
\begin{itemize}
\item Introduce an inner core, made of a static perfect fluid, as shown in Fig.\ref{f9} left. The core represents a compressed matter, to which RDM flows are connected on the surface.
\item Apply the model of relativistic networks again, as shown in Fig.\ref{f9} right. The core is resolved as a randomized network, consisting of normal and/or tachyonic particles, connected in the interaction vertices. The model \cite{cosmic-web} should be adjusted to the curved (static spherically symmetric) space-time.
\item Connect the interior of RDM model with a wormhole, leading to a different domain of the space-time. The radially converging flows of dark matter become radially diverging on the other side. The models replacing central black holes in the galaxies by wormholes are becoming widespread \cite{wormhole1,wormhole2,wormhole3}.
\end{itemize}
While the present study focuses mainly on the question, what happens if the dark matter flows radially diverge from the center of the galaxy, the proposed extensions can answer the question, why they are doing so. The base model assumes that at normal energies the dark matter is passive and on the level of elementary particles does not interact with itself and other types of matter, except of the gravitational way. The first two extensions assume that at high energies, available only in the inner core, the dark matter starts to interact with itself and with the matter in the core. In this way the world lines of dark matter become connected to the core, i.e., can start and end only in the core. Since the black holes become sources and sinks of dark matter, at large distances the world lines of dark matter radially diverge from the black holes. The third possibility considers the world lines passing through the wormhole and then radially diverging from it. In this case the dark matter can be completely non-interactive, while its world lines become connected to the wormholes by the topology of the space-time.

\section{Conclusion}\label{concl}
We have studied a static spherically symmetric problem with a black hole and $T$-symmetric radially directed geodesic flows of dark matter. The problem is relevant to the modeling of the galaxies. Its solutions have the following structure. At large distances the gravitational acceleration decreases with a distance as $\sim r^{-1}$, corresponding to flat rotation curves of the galaxies. At smaller distances the dependence switches to Newtonian regime $\sim r^{-2}$, then to Schwarzschild regime. The deviation from Schwarzschild regime starts below the gravitational radius $r_s$, where the dark matter prevents the creation of event horizon. Instead, a region is created where the metric components $g_{tt}<0$ and $g_{rr}>0$ are extremely small. Although the region possesses extremely large redshift and for an external observer looks completely black, the observer can enter the region and leave it in a finite time. The structure of radial geodesics implies that the solution has a single patch, where standard $(t,r)$ coordinates are globally applicable. The analysis of circular geodesics shows the second island of stable circular orbits present under the gravitational radius. Matching of the model with parameters of the Milky Way galaxy allows to reconstruct the ranges of typical field behavior. Below the gravitational radius the model shows principally different structure of the central black hole. At large distances the model reproduces observable asymptotically flat rotation curves, providing for them a natural explanation, which does not require a special adjustment of parameters. The model also predicts the range, where the rotation curves have almost Keplerian behavior, containing the observed orbits of S0-2 and S0-102 stars. The possible extensions of the model have been discussed, including connections to the models of filaments, a hot inner core composed of interacting dark matter and a wormhole directing the dark matter to the other part of the universe.

\end{document}